\documentclass[]{article}
\usepackage{graphicx}
\usepackage{amsmath}
\usepackage{xcolor}
\usepackage{authblk}
\usepackage{hyperref}
\usepackage{amssymb}
\usepackage{mathtools}
\usepackage{array}
\usepackage{float}
\usepackage{caption}
\usepackage{subcaption}

\newcommand{\Zstroke}{%
	\text{\ooalign{\hidewidth\raisebox{0.2ex}{--}\hidewidth\cr$Z$\cr}}%
}
\newcommand{\zstroke}{%
	\text{\ooalign{\hidewidth -\kern-.3em-\hidewidth\cr$z$\cr}}%
}
\title{Running vacuum model versus $\Lambda$CDM - a Bayesian analysis}
\author{Sarath N, \footnote{\href{mailto:sarath@cusat.ac.in}{sarath@cusat.ac.in}} and 
	Titus K. Mathew,  \footnote{\href{mailto:titus@cusat.ac.in}{titus@cusat.ac.in}} 
}
\affil{\textit{\small Department of Physics, Cochin University of Science and Technology, \\ Kochi, Kerala 682022, India}}
\date{}
\begin{document}
	
	\maketitle
	\begin{abstract}
		We study the running vacuum model in which the vaccum energy density depends on square of Hubble parameter in comparison with the $\Lambda$CDM model. In this work, the Bayesian inference method is employed to test against the standard $\Lambda$CDM model to appraise the relative significance of our model, using the combined data sets, Pantheon+CMB+BAO and Pantheon+CMB+BAO+Hubble data. The model parameters and the corresponding errors are estimated from the marginal likelihood function of the model parameters. Marginalizing over all model parameters with suitable prior,  we have obtained the Bayes factor as the ratio of Bayesian evidence of our model and the $\Lambda$CDM model. The analysis based on Jeffrey's scale of bayesian inference shows that the evidence of our model against the $\Lambda$CDM model is weak for both data combinations. Even though the running vacuum model gives a good account of the evolution of the universe, it is not superior to the $\Lambda$CDM model. 
	\end{abstract}
\section{Introduction}

A comprehensive probe of the neoteric cosmological observations shows that our Universe is spatially flat on a large scale and composed of baryonic matter ($\sim 4\%$), dark matter ($\sim 26\%$) and dark energy ($\sim 70\%$) \cite{Papagiannopoulos2020}. The observation that the Universe is undergoing an accelerated expansion is explained using the exotic form of energy with a negative pressure called dark energy \cite{frieman2008dark, bjorken2003cosmology}. This exotic form is evident from the cosmological observations; the emergence and nature remain a mystery \cite{astier2012observational}. 
The concordance model of cosmology or $\Lambda$CDM model incorporates cold dark matter with dark energy to explain the recent accelerated expansion of the Universe in the light of supernovae data. It also explain the existence and structure of Cosmic microwave background radiation (CMB), the large-scale structure in the distribution of galaxies and the observed abundances of hydrogen, helium, and lithium \cite{peebles2003cosmological}. The model assumes a rigid cosmological constant as a candidate for dark energy. Apart from its splendid prediction power, it has some downsides; the cosmological constant problem: experimentally observed value of the cosmological constant density is many orders of magnitude less as compared to the value predicted by the standard model of particle physics, the coincidence problem: the coincidence of the present value of the cosmological constant density and the dark matter density \cite{Sola:2015rra}. The recent development in this field suggests that dynamical vacuum energy is a suitable candidate for dark energy \cite{sola2015fundamental, fritzsch2017running, freese1987cosmology, alam2004case}. The running vacuum model in which vacuum energy density depends on the Hubble parameter as $\rho_{\Lambda}(H)\sim H^2$ is a suitable choice. This perception has got much attention in the light of recent cosmological data. The Vacuum energy acquires its time dependence by the time dependence of the Hubble parameter, and that is implied by the renormalization group approach of quantum field theory in curved spacetime \cite{Sola:2015rra}. Such a model with the equation state $\omega_{\Lambda}=-1$ can resolve the cosmological constant problem and the coincidence problem. Recent combined observational data of SNIa+BAO+H(z)+LSS+BBN+CMB report strong evidence of a slowly decaying cosmological constant \cite{sola2017first}. With the potential ability of the model to explain the recent acceleration of the Universe and its experimental supports\cite{zhao2017dynamical}, it is a worthy exercise to compare the running vacuum model with the standard $\Lambda$CDM model based on the Bayesian statistics.\\ \vspace{0cm}
\hspace{0.25cm}We compare two cosmological models by adopting available statistical methods as a large variety of cosmological data is available. We have two distinct approaches in statistics, the frequentist and the Bayesian statistics. Both these methods allow one to obtain evidence out of competing hypotheses. The former one is not much useful in cosmology as creating ensembles of the Universe is not possible. On the other hand, Bayesian statistics found its place in cosmology as it interprets probability as a degree of belief rather than ensembling \cite{padilla2019cosmological}. Based on conditional probability, Bayesian statistics was introduced by Thomas Bayes and advanced by great mathematicians such as Gauss, Bayes, Laplace, Bernoulli \cite{john2002comparison, george2020bayesian}. In cosmology, the possibility of assigning a probability to a random variable is not possible or rarely possible because ensembles and repeated measurements are hardly possible. Instead, we use a Hypothesis or model that can either be appropriate or inappropriate \cite{hobson2010bayesian}. The Bayesian statistics allow us to assign a probability to the model based on the data available. This method gives us an excellent description to obtain the marginal likelihood of the model parameters and differentiate two models by calculating the Bayes factor, which is, by definition, the ratio of the Bayesian evidences \cite{verde2010statistical}. In this work, we compare the running vacuum model with the standard $\Lambda$CDM model using Bayesian inference.

The paper is organized in the following manner. In section 2, we describe the Bayesian analysis strategy. Section 3 gives a detailed description of the running vacuum model.   In section4, we perform Bayesian inference to extract the Bayes factor by assuming appropriate priors. Finally, conclude in the last section.
\section{Bayesian analysis strategy}
Bayesian statistics is an essential mathematical tool that found its place in cosmology to estimate a combination of model parameters that best describe the Universe and the model comparison \cite{Padilla:2019mgi}.  The method is based on the view of probability as credence rather than an ensemble. The cornerstone of Bayesian statistics is the Bayes theorem; it is a direct consequence of axioms of probability. It provides us with a gratifying description to figure out the posterior probability, $P(\theta|D, \mathcal{H})$, the probability of the existence of the parameter vector of the model in the light of observational data (D) and for a given model or hypothesis ($\mathcal(H)$). According to Bayes theorem,
\begin{equation}
\label{eqn:p1}
P(\theta|D,\mathcal{H})= \frac{P(\theta|\mathcal{H})P(D|\theta,\mathcal{H})}{P(D|\mathcal{H})}.
\end{equation} 
Here, $P(\theta|\mathcal{H})\equiv \pi(\theta)$ is the prior probability which convey any information about the model before acquiring data. There is no predefined prescription to choose suitable prior for the analysis; rather, it solely depends on the researcher's allied knowledge and experience in the field and the quality of judgment. However, once a prior has been selected, the iterated application of Bayes theorem leads to convergence to a common posterior\cite{john2002comparison}. It is important to specify the prior explicitly in Bayesian analysis; otherwise, readers may not be able to reproduce the result. The term  $P(D|\theta, \mathcal{H})\equiv \mathcal{L}(D|\theta,\mathcal{H})$ is the likelihood function, or simply the likelihood, which defines the probability of getting data, given the model is true for a given set of parameters. The term $P(D|\mathcal{H})\equiv \Zstroke$  is just a normalization factor that defines the evidence of the model, frequently called Bayesian evidence. It is solely the average of the likelihood over the prior for a specific model of choice. 
\begin{equation}
\label{eqn:p01}
P(D|H) = \int d^N\theta P(D|\theta,\mathcal{H})P(\theta|\mathcal{H}),
\end{equation} 
where N is the dimension of the parameter space. When we are dealing with parameter space of a unique model, this quantity can be avoided. But the Bayesian evidence plays a key role while performing the model selection\cite{Padilla:2019mgi}. 

In the Bayesian theory of model selection, we compare the two models under consideration by evaluating odds or Bayes factor, is given by,
\begin{equation}
\label{eqn:p02}
B_{ij} \equiv \frac{P(D|\mathcal{H}_i)}{P(D|\mathcal{H}_j)}=\frac{\int d^{N_i}\theta_i \mathcal{L}(D|\theta_i,\mathcal{H}_i)\pi(\theta_i)}{\int d^{N_j}\theta_j \mathcal{L}(D|\theta_j,\mathcal{H}_j)\pi(\theta_j)} =\frac{\Zstroke_i}{\Zstroke_j},
\end{equation}  
where $\theta_i$ is a parameter vector for the hypothesis $\mathcal{H}_i$. The Bayes factor gives us a better understanding on how well the model $\mathcal{H}_i$ match the observational data when compared to model $\mathcal{H}_j$. The likelihood function $\mathcal{L}(D|\theta_i,\mathcal{H}_i)$ can be estimated using the expression,
\begin{equation}
\label{eqn:l2}
\mathcal{L}(D|\theta_i,\mathcal{H}_i) \equiv exp(-\chi^2(\theta_i)/2),
\end{equation} 
where we assumed the measurement errors are Gaussian. The $\chi^2$ can be evaluated using the expression,
\begin{equation}
\label{eqn:chi}
\chi^2(\theta_i) = \sum \left[\frac{A_k - A_k(\theta_i)}{\sigma_k}\right]^2.
\end{equation} 
Here, $A_k$ is the value obtained from the data sets, $A_k(\theta_i)$ is the corresponding theoretical value obtained from our model and $\sigma_k$ is the error in the measured values. We choose uniform prior for the model parameters that lie in the interval $[\theta_i,\theta_i+\Delta\theta_i]$ such that the prior probability of the model parameters become $\pi(\theta_i) = \frac{1}{\Delta\theta_i}$. Then, the equation (\ref{eqn:p01}) can be re-written as,
\begin{equation}
\label{eqn:l3}
\Zstroke_i = \frac{1}{\Delta\theta^1_i...\Delta\theta^N_i}\int_{N}d{\theta^1_i}'...d{\theta^N_i}'\exp[-\chi^2({\theta^1_i}'...{\theta^N_i}')/2].
\end{equation} 
\begin{table}
	\begin{tabular*}{\columnwidth}{@{}l@{\hspace*{30pt}}l@{\hspace*{40pt}}l@{}}
		\hline
		Bayes Factor & Comment \\ \hline
		$B_{ij}<1$ & $M_i$ is not significant as $M_j$ \\ \hline
		$1<B_{ij}<3$ & Evidence of $M_{i}$ against $M_{j}$ is weak \\ \hline
		$3<B_{ij}<20$ & Evidence of $M_i$ against $M_j$ is definite \\ \hline
		$20<B_{ij}<150$ & Evidence of $M_i$ against $M_j$ is strong \\ \hline
		$B_{ij}>150$ & Evidence of $M_i$ against $M_j$ is very strong \\ \hline
	\end{tabular*}
\caption{Jeffreys scale of Bayesian inference}
\label{tab:anysymbols}
\end{table}
As the model $\mathcal{H}_i$ depends on more than one parameter, we use marginalization to obtain the posterior probability distribution of the parameter of interest. It is also possible to find all the model parameters that best fit the data, freeze all the model parameters except the parameter of interest to its best fit value and vary the parameter of interest to find its posterior probability distribution. Nevertheless, this procedure is incorrect as it yields correct results only in special circumstances. Here, the model $\mathcal{H}_i$ has N independent parameters, $\theta_i^1$, $\theta_i^2$, $\theta_i^3$....$\theta_i^N$. To obtain the posterior probability distribution of the parameter of interest, say $\theta_i^1$, marginalize over all other parameters by
\begin{equation}
\label{eqn:l4}
P(\theta_i^1|D,\mathcal{H}_i) = \frac{1}{\Delta\theta^2_i...\Delta\theta^N_i}\int_{N-1}d{\theta^2_i}'...d{\theta^N_i}'\exp[-\chi^2({\theta^2_i}'...{\theta^N_i}')/2],
\end{equation}
where N is the total number of parameters in the model $\mathcal{H}_i$.

We use conventional Jeffreys scale of inference for the analysis purpose \cite{Nesseris:2012cq, george2020bayesian}, which is presented in Table \ref{tab:anysymbols}.
\section{Running Vacuum Model}
According to general theory of relativity, the geometric structure of region of space-time is not self-reliant but determined by the energy-momentum tensor. For a universe with a perfect fluid having energy density $\rho$ and pressure $p=\omega\rho$, where $\omega$ is the equation of state of the fluid, the energy- momentum tensor is given by $T_{\mu\nu} = -pg_{\mu\nu} + (\rho + p)u_{\mu}u_{\nu}$, where $g_{\mu\nu}$ is the metric tensor and $u_{\mu}$ is the four-velocity. In dynamical vacuum cosmology, we consider an additional term $g_{\mu\nu}\rho_v(H(t))$, $\rho_v(H(t)) = \frac{\Lambda(H(t))}{8\pi G}$ with the energy momentum tensor. Then, the effective energy-momentum tensor can be written as $\tilde{T}_{\mu\nu} \equiv T_{\mu\nu} + g_{\mu\nu}\rho_v(H)$, where $H = \frac{\dot{a}}{a}$ is the Hubble parameter. In this framework, the Einstein field equation can be expressed as,
\begin{equation}
\label{eqn:E1}
R_{\mu\nu}-\frac{1}{2}g_{\mu\nu}R = 8\pi G\tilde{T}_{\mu\nu}.
\end{equation} 
The Friedmann equations that govern the expansion of the universe with a spatially flat FLRW metric $ds^2 = -dt^2 + a^2(t)(dx^2 + dy^2 + dz^2)$, where a is the scale factor, are
\begin{equation}
\begin{aligned}
\label{eqn:F01}
&3H^2 = 8\pi G(\rho_m + \rho_{\gamma} + \rho_{v})\\
&2\dot{H} + 3H^2 = -8\pi G(\Omega_{m_0}\rho_m + \omega_{\gamma}\rho_{\gamma} + \omega_{v}\rho_{v}),
\end{aligned}
\end{equation}
where $\Omega_{m_0}$, $\omega_{\gamma}$, $\omega_{v}$ are equation of state for matter, radiation and vacuum energy respectively.
In this work, we consider matter-late accelerating epoch of the Universe. Considering the fact that the radiation doesn't contribute significantly in the matter-late accelerating epoch, and non relativistic matter with the equation of state $\Omega_{m_0} = 0$, and vacuum energy with equation of state $\omega_{\Lambda} = -1$, we obtain the Friedmmann equation in a reduced form as,
\begin{align}
\label{eqn:F1}
&3H^2 = 8\pi G(\rho_m + \rho_{v}),\\
\label{eqn:F2}
&2\dot{H} + 3H^2 = 8\pi G\rho_{v}.
\end{align}
Note that the $\rho_v$ depends on time, and it acquires its time dependence from the Hubble parameter.  It is possible to obtain the generalized conservation law in the $\Lambda$-varying framework by using the explicit form of the Friedmann-Lema$\hat{i}$tre-Robertson (FLRW) metric. The explicit form in the matter and vacuum energy dominated Universe takes the form
\begin{equation}
\label{eqn:c1}
\dot{\rho}_m + 3H\rho_m = -\dot{\rho}_v.
\end{equation} 
Here we consider the running vacuum model; a dynamical dark energy model developed based on the expectation that an expanding universe may not have a static vacuum energy density \cite{Sola:2015rra}. This idea is theoretically motivated by the renormalization group approach of quantum field theory in curved spacetime. We associate the renormalization Group's running scale $\mu$ with the energy threshold associated with the cosmology scale \cite {geng2020constraints}. Thus $\mu$ can be chosen to be the Hubble expansion rate $H$ that defines the expansion of the Universe, which allows us to express vacuum energy density as a power series of the Hubble expansion rate.   
\begin{equation}
\label{eqn:rho01}
\rho_{v}(H) = \frac{3}{8\pi G}\left(c_0 + \sum_{k}\alpha_k H^k\right).
\end{equation} 
The general covariance of effective action of quantum field theory in curved spacetime allows only the even powers of Hubble parameter. Moreover, the higher-order terms of the Hubble parameter do not contribute significantly in the matter--late accelerating Universe (Note that the high powers of $H$ are beneficial to explain the early Universe) \cite{Papagiannopoulos2020}. Therefore, the vacuum energy density takes the form,
\begin{equation}
\label{eqn:rho1}
\rho_{v}(H) = \frac{3}{8\pi G}\left(c_0 + \nu H^2\right).
\end{equation} 
Here, $c_0$ plays the role of the cosmological constant and $\nu$ is a dimensionless constant that plays a role similar to the $\beta$--function coefficient within the structure of effective action in quantum field theory in curved space time \cite{lima2013expansion}. The coefficient $\nu$ depends on the square of the masses of matter particles, can be written as \cite{Sola:2015rra},
\begin{equation}
\label{eqn:nu}
\nu = \frac{1}{6\pi}\sum_{i=f,b}^{}B_i\frac{M_i^2}{M_P^2}.
\end{equation} 
These models come up with a better explanation of the cosmological observations as compared to the standard $\Lambda$CDM model. The slowly varying cosmological constant is supported by observational data on type Ia supernovae (SN1a), the Cosmic Microwave Background (CMB), and Baryonic Acoustic Oscillations (BAO). The model is having the running vacuum energy and dark matter as the components that determine the evolution of the Universe. 
Combining equation (\ref{eqn:F1}) and (\ref{eqn:F2}), we obtain
\begin{equation}
\label{eqn:H1}
\dot{H} = -4\pi G\rho_m.
\end{equation}
Substituting equation (\ref{eqn:rho1}) and (\ref{eqn:H1}) in (\ref{eqn:c1}), we obtain the conservation equation of matter,
\begin{equation}
\label{eqn:rho3}
\dot{\rho_m} + 3(1-\nu)H\rho_m = 0.
\end{equation}
In order to solve the equation, change integration variable from time to scale factor, we arrive at
\begin{equation}
\label{eqn:rho4}
\rho_{m}(a) = \rho_{m_0}a^{-3(1-\nu)},
\end{equation}
where $\rho_{m_0}$ is the present value of the matter density. Substituting equation (\ref{eqn:rho4}) in (\ref{eqn:H1}), we obtain the evolution of the Hubble parameter that describe the evolution of the Universe in the matter-late accelerating epoch 
\begin{equation}
\label{eqn:H3}
H(z) = H_0\sqrt{1+\frac{\Omega^0_m}{1-\nu}\left[\left(1+z\right)^{3(1-\nu)}-1\right]},
\end{equation}
where $z = \frac{1-a}{a}$ is the redshift, $H_0$ is the Hubble parameter at present and $\Omega_{m_0} = \frac{8\pi G\rho_{m_0}}{3H_0^2}$ is the present value of the matter density.
The vacuum energy density can be obtained by substituting equation (\ref{eqn:rho4}) and (\ref{eqn:H3}) in equation (\ref{eqn:F1}),
\begin{equation}
\label{eqn:rho2}
\rho_{v}(a) = \rho_{v_0} + \frac{\nu}{1-\nu}\rho_{m_0}\left(a^{-3(1-\nu)}-1\right),
\end{equation}
where $\rho_{v_0} = \frac{3H_0^2}{8\pi G}-\rho_{m_0}$ is the present value of the vacuum energy density. The equation (\ref{eqn:H3}) represents the model or the hypothesis, that can be used to test against $\Lambda$CDM model to obtain the relative significance of our model in the light of observational data. 
\section{Bayesian inference}
The running vacuum model ($M_{RVM}$) possess $H_0$, $\Omega_{m_0}$, $\Omega_{\Lambda_0}$ and $\nu$ as the free parameters. The matter density parameter $\Omega_{m_0}$ and vacuum density parameter $\Omega_{\Lambda_0}$ are related by the constraint $\Omega_{m_0} + \Omega_{\Lambda_0} = 1$. Hence the number of free parameters reduces to three. Now, we have to extract these parameters that best describe the universe using observational data. We adopt parameter inference procedure to estimate the marginal likelihood of all the model parameters using the data combinations Panthoen+CMB+BAO and Panthoen+CMB+BAO+Hubble data. The pantheon data set hold 1048 SNIa data in the redshift range $0 < z < 2.3$ \cite{scolnic2018complete, wang2021exploring}. We have used shift parameter ($\mathcal{R}$) of the Cosmmic Microwave Background (CMB) data from Planck2018, the acoustic parameter ($\mathcal{A}$) of Baryonic Acoustic Oscillations (BAO) data from SDSS \cite{eisenstein2005detection,tegmark2006cosmological,beutler20116df} and the Hubble parameter data holding 38 data in the red shift  span $0.07\leq z \leq 2.36$ \cite{farooq2017hubble}. The marginal likelihood of a particular model parameter, say $\nu$ can be obtained using equation (\ref{eqn:l4}), where we have to integrate over other two parameters $H_0$ and $\Omega_{m_0}$, that will give rise to,
\begin{align}
\label{eqn:l5}
P(\nu|D,M_{RVM}) = \frac{1}{\Delta H_0} \frac{1}{\Delta\Omega_{m_0}}\int_{H_0}^{H_0+\Delta H_0}\int_{\Omega_{m_0}}^{\Omega_{m_0}+\Delta\Omega_{m_0}}\nonumber \\d H_0' d\Omega_{m_0}'
\exp[-\chi^2(\nu, H_0',\Omega_{m_0}')/2].
\end{align}
Similarly, we can evaluate the marginal likelihood of other parameters too. To obtain this, primarily we have to estimate the $\chi^2$ using equation (\ref{eqn:chi}) for the data combinations Pantheon+CMB+BAO and Pantheon+CMB+BAO+Hubble data. To obtain $\chi^2$, we use equation (\ref{eqn:chi}), where we replace $A_k$ with the observed physical quantity from the data and $A_k (\theta_i)$ with the corresponding theoretical one. In Pantheon data, the observation gives the apparent magnitude of the Type 1a supernovae. The corresponding theoretical one can be calculated using the expression,
\begin{align}
\label{eqn:m1}
m (H_0, \Omega_{m_0}, \nu, z_i) = 5\log_{10} \left[\frac{d_L(H_0, \Omega_{m_0}, \nu, z_i)}{Mpc}\right]\nonumber \\+25+M
\end{align}
where M is the absolute magnitude of the Type 1a Supernovae. Its magnitude can be evaluated using chi square minimization, the best fit value is $-19.35$ for Pantheon+CMB+BAO and $-19.40$ for Pantheon+CMB+BAO + Hubble data. The luminosity distance $d_L$ is related to Hubble parameter as,
\begin{align}
\label{eqn:m2}
d_L = c(1+z)\int_{0}^{z}\frac{dz'}{H},
\end{align}
where $c$ is the speed of light in vacuum.
To obtain chi square  using Hubble data, We replace $A_k$ with $H$ from  Hubble data and $A_k (\theta_i)$ with theoretical Hubble parameter given in equation (\ref{eqn:H3}). To obtain $\chi^2$ using CMB data,  we use shift parameter $\mathcal{R}$ as the observable in lieu of $A_k$ and $A_k(\theta_i)$ is replaced with theoretical shift parameter that is obtained using the equation \cite{elgaroy2007using},
\begin{align}
\label{eqn:shift}
\mathcal{R} = \sqrt{\Omega_{m_0}}\int_{0}^{z_2}\frac{dz}{h(z)}
\end{align}
where $z_2$ is the redshift at the cosmic photosphere and $h(z) = \frac{H(z)}{H_0}$ is the reduced Hubble parameter. From Planck 2018 data, $z_2 = 1089.92$ and the shift parameter, $\mathcal{R} = 1.7502\pm0.0046$. To obtain the $\chi^2$ using BAO data, the acoustic parameter $\mathcal{A}$ is used as the observable instead of $A_k$, the theoretical acoustic parameter can be estimated using the equation\cite{ryan2018constraints},
\begin{align}
\label{eqn:shift}
\mathcal{A} = \frac{\sqrt{\Omega_{m_0}}}{h(z_1)^\frac{1}{3}}\left(\frac{1}{z_1}\int_{0}^{z_1}\frac{dz}{h(z)}\right)^\frac{2}{3},
\end{align}
where $z_1$ is the redshift corresponding to which the signature of peak acoustic oscillation has been measured. The estimated value of shift parameter with reference to the SDSS data for $z_1 = 0.35$ is, $\mathcal{A} = 0.484\pm0.016$ \cite{blake2011wigglez}. 

\begin{figure}[H]
	\centering
	\begin{subfigure}{0.45\linewidth}
		\centering
	\includegraphics[width=1\textwidth]{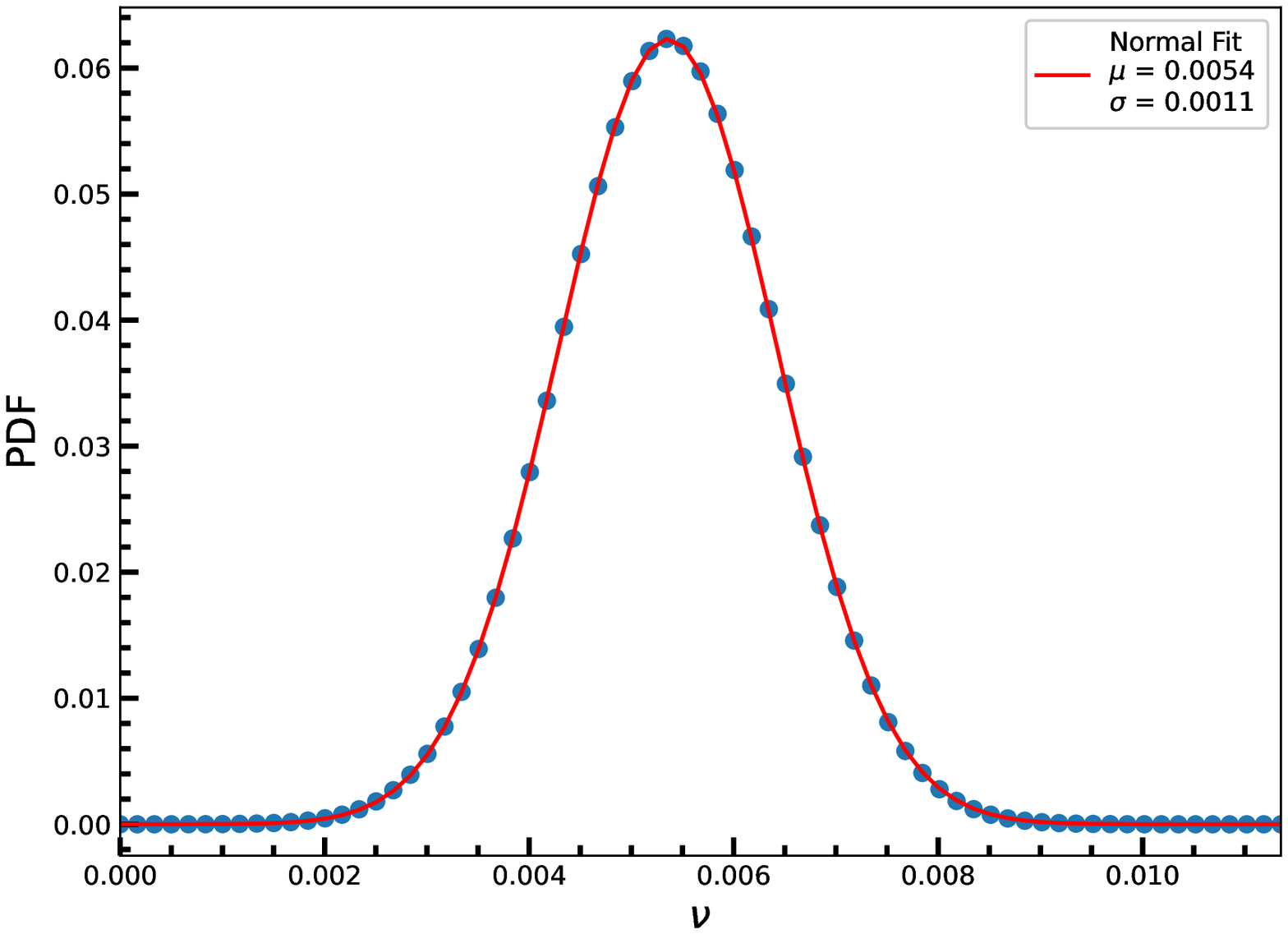}
	\caption{Data1}
	\label{fig:nu1}
\end{subfigure}
\begin{subfigure}{0.45\linewidth}
	\centering
	\includegraphics[width=1\textwidth]{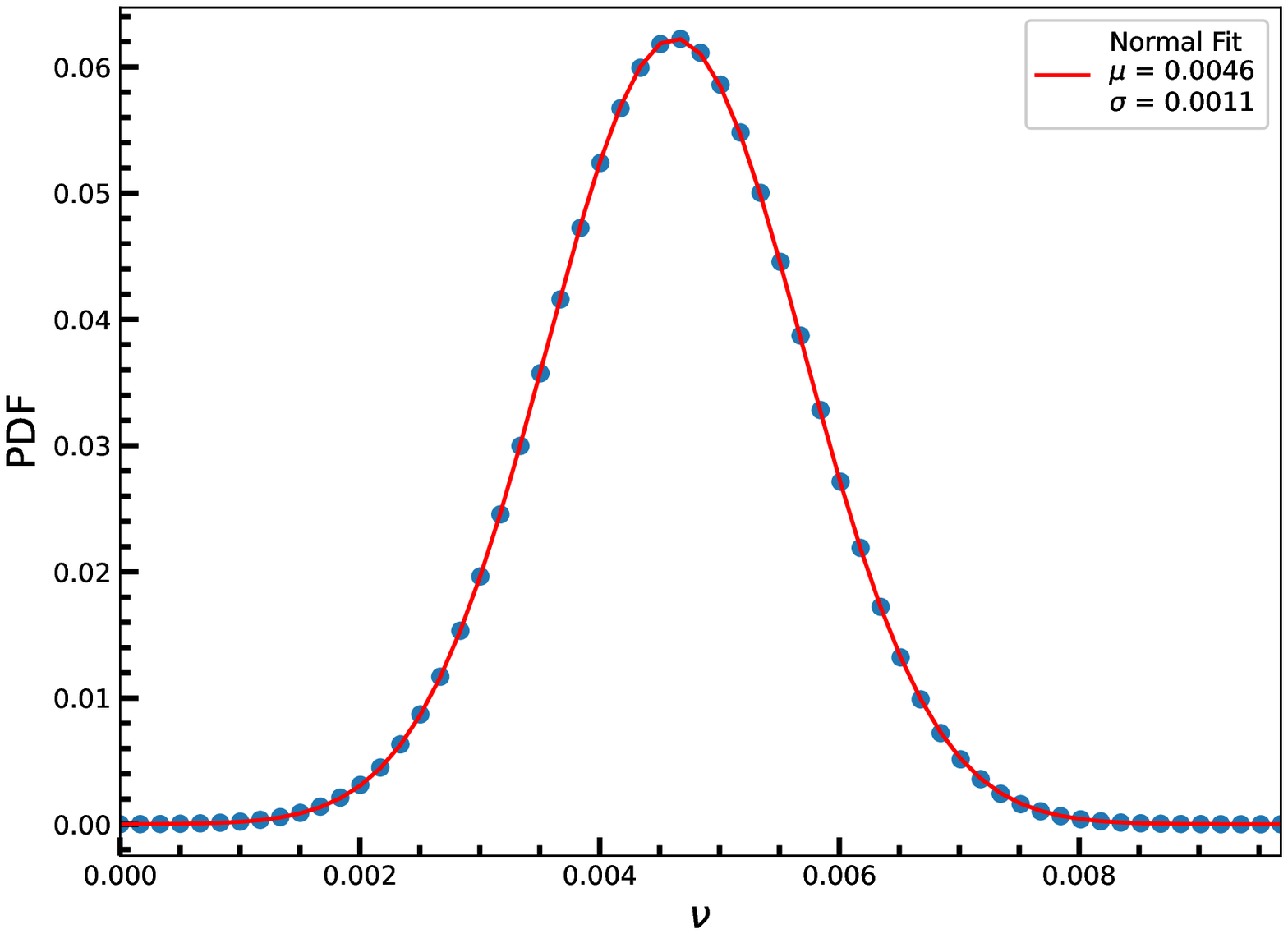}
	\caption{Data2}
	\label{fig:nu2}
\end{subfigure}
\caption{Probability density function of $\nu$ is plotted fitted with Gaussian.}
\end{figure}

\begin{figure}[H]
	\centering
	\begin{subfigure}{0.45\linewidth}
		\centering
		\includegraphics[width=1\textwidth]{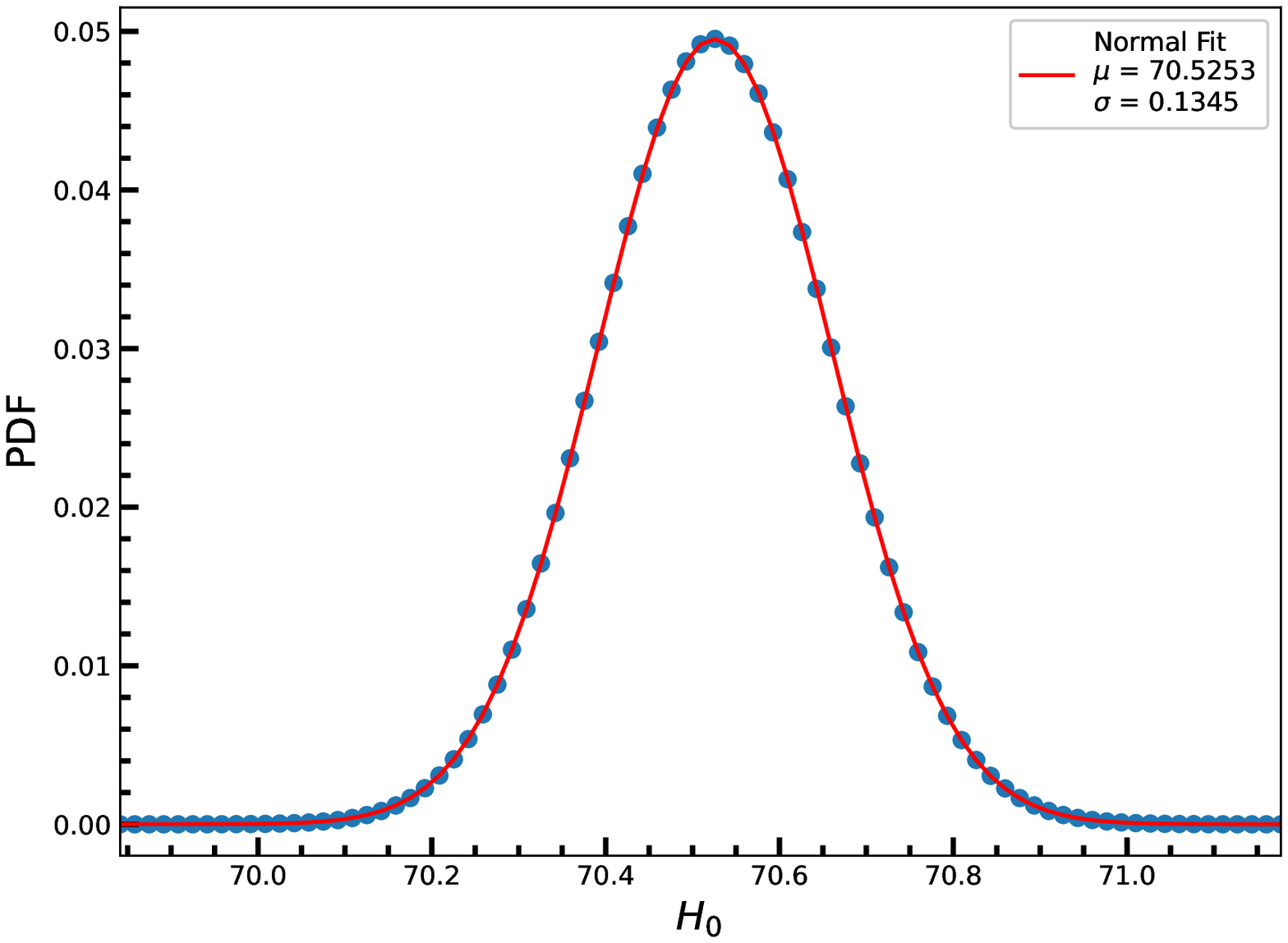}
		\caption{Data1}
		\label{fig:H1}
	\end{subfigure}
	\begin{subfigure}{0.45\linewidth}
		\centering
		\includegraphics[width=1\textwidth]{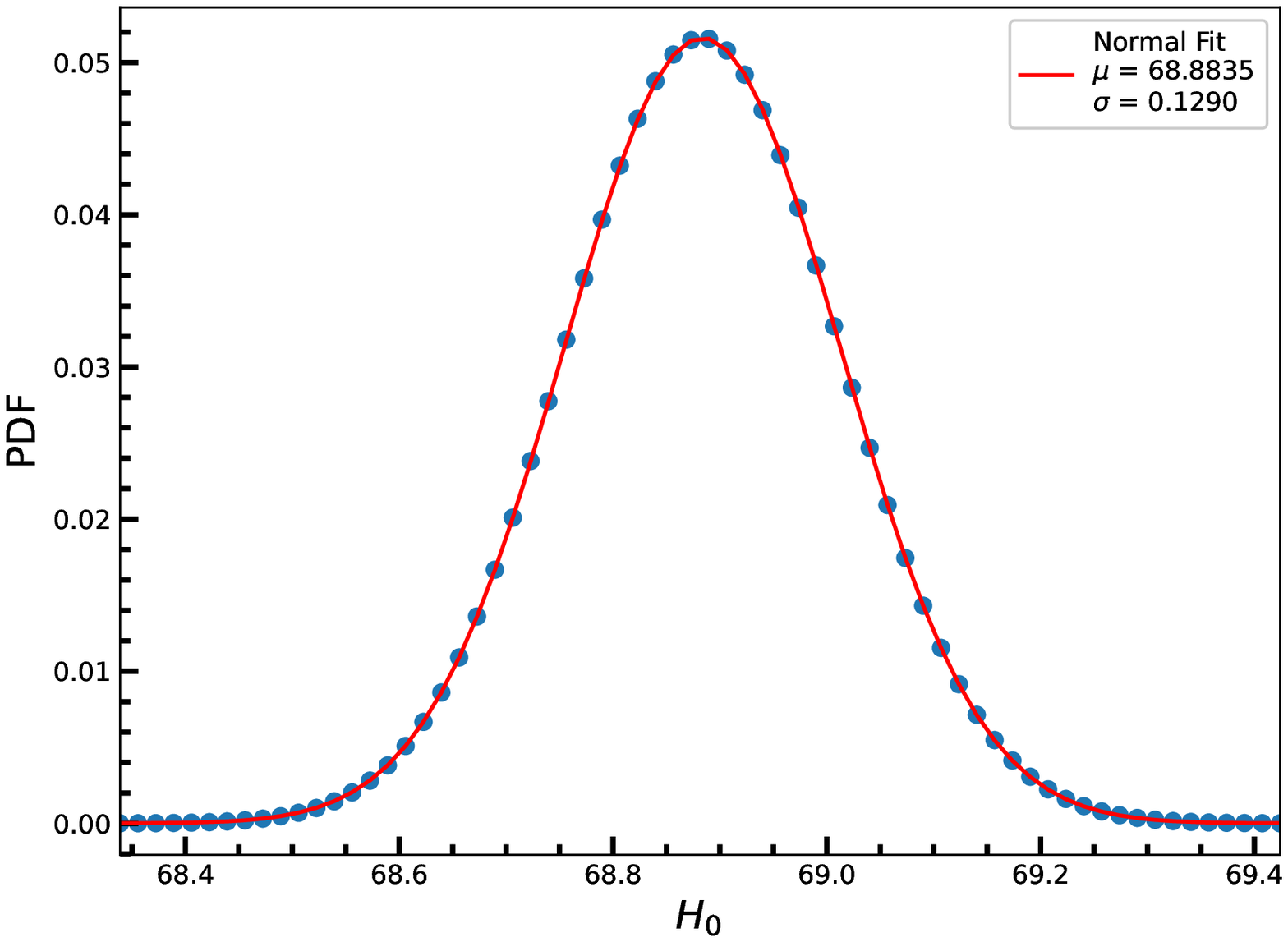}
		\caption{Data2}
		\label{fig:H2}
	\end{subfigure}
	\caption{Probability density function of $H_0$ is plotted fitted with Gaussian.}
\end{figure}

\begin{figure}[H]
	\centering
	\begin{subfigure}{0.45\linewidth}
		\centering
		\includegraphics[width=1\textwidth]{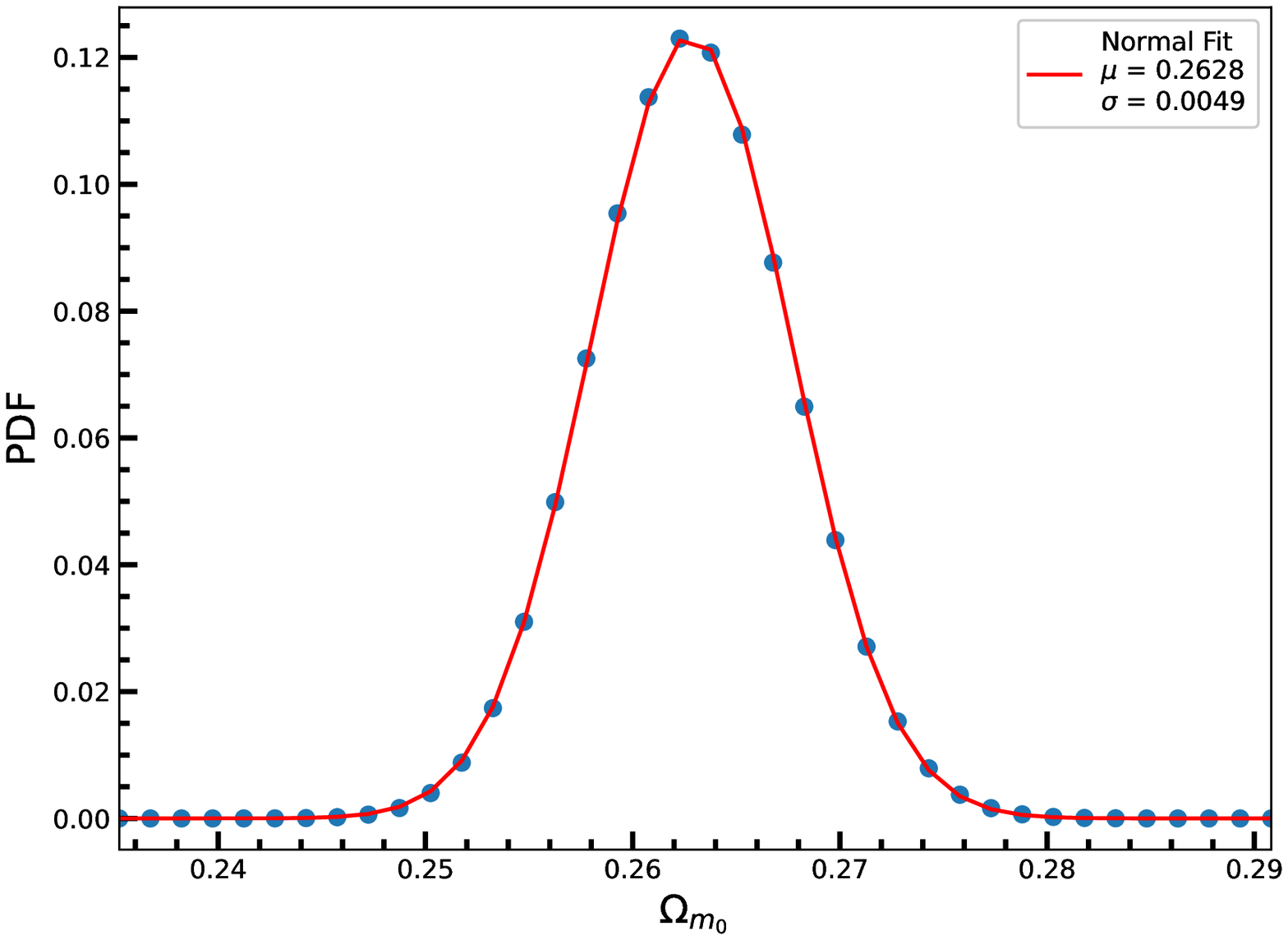}
		\caption{Data1}
		\label{fig:om1}
	\end{subfigure}
	\begin{subfigure}{0.45\linewidth}
		\centering
		\includegraphics[width=1\textwidth]{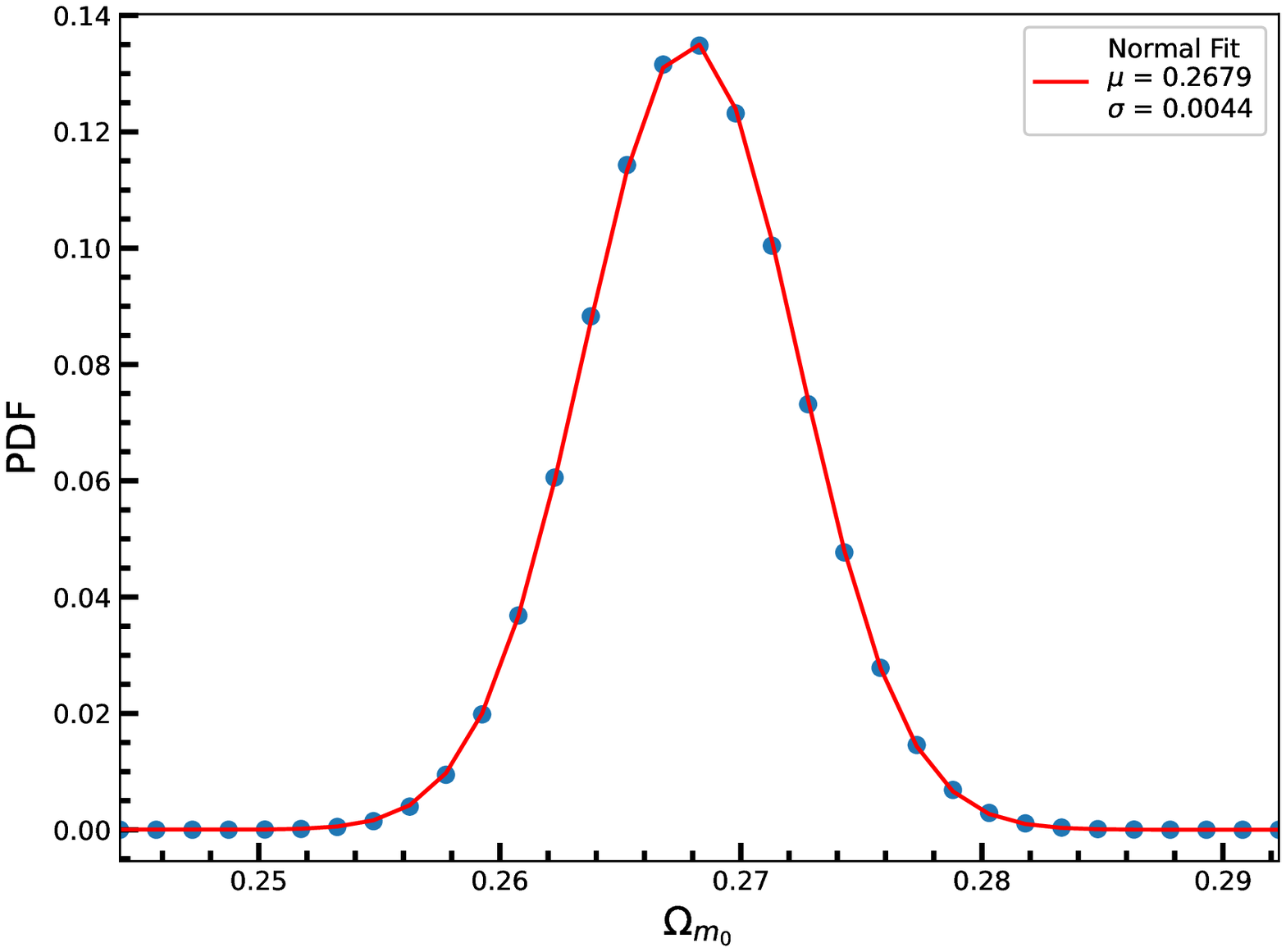}
		\caption{Data2}
		\label{fig:om2}
	\end{subfigure}
	\caption{Probability density function of $\Omega_{m_0}$ is plotted fitted with Gaussian.}
\end{figure}

We have evaluated the chi-square for the data combinations Pantheon+CMB\\+BAO  and Pantheon+CMB+BAO+Hubble data which give rise to a combined $\chi^2$ of the form, $\chi^2=\chi^2_{pantheon}+\chi^2_{CMB}+\chi^2_{BAO}$ and $\chi^2=\chi^2(H_0,\Omega_{m_0},\nu)_{pantheon}+\chi^2_{CMB}+\chi^2_{BAO} +\chi^2_{Hubbledata}$ respectively. We carried out parameter inference by calculating the marginal likelihood of the model parameters $H_0, \Omega_{m_0}, \nu$. The marginal likelihood of the parameters was obtained using equation (\ref{eqn:l5}). To evaluate marginal likelihood, it is necessary to specify the priors. We choose flat priors for all the model parameters. We endorse a prior for $\nu$ in the range $0 - 0.1$. As discussed earlier, the parameter $\nu$ is the coefficient of $H^2$ in the expression of vacuum energy density, and it provides the running status to the vacuum energy. The parameter $\nu =0$ represents the constant vacuum energy density, and that corresponds to the standard $\Lambda$CDM model. The term $\nu H^2$ represent a small variation from the cosmological constant, the parameter $\nu\ll1$ \cite{Sola:2015rra}. Since $\nu$ is directly related to the ratio of the weighted sum of squares of all masses contributing the loop and the square of the Planck mass \cite{sola2008dark}, it cannot be negative. The non-negative value of $\nu$ keeps the vacuum energy density strictly positive. The typical value obtained from the fit of combined data on SNIa, the shift parameter of CMB, and BAO data is $\nu=10^{-3}$. The parameter range $0<\nu<2.83\times10^{-4}$ is obtained by the analysis of CMB power spectrum and baryon acoustic oscillation data \cite{geng2020running}. This vindicate the range of $\nu$ as $0 - 0.1$. We choose a uniform prior for the matter density parameter $\Omega_{m_0}$ in the range $0-1$. A uniform prior of $0.001$ to $0.99$ is assumed for the Bayesian analysis of viscous dark energy model \cite{da2019extended}. A uniform prior within the range $0\leq \Omega_{m_0}\leq0.8$ is used in the Ref. \cite{gupta2010direction}. The constraint on $\Omega_{m_0}$ is restricted to $0\leq\Omega_{m_0}\leq1.0$ to explain supernovae data in the framework of pure cold dark matter model \cite{lima2011deformed}. We adopt a parameter range for $H_0$ in the range $65-75$ $kms^{-1}Mpc^{-1}$. The recent observation by Planck collaboration (2018) measured value of $H_0$ as $67.4\pm0.5$ $km s^{-1}Mpc^{-1}$ \cite{aghanim2020planck}. The best estimate value of $H_0$ obtained from the local expansion rate is $73.24\pm1.74$ $kms^{-1}Mpc^{-1}$ \cite{riess20162}.
The marginal Probability Density Function of the model parameter $\nu$ is obtained by integrating over all the model parameters except $\nu$ using equation (\ref{eqn:l5}).
The probability density function is fitted with the Gaussian function to obtain the most probable value and the error given in Table \ref{tab:table2}. 
\begin{table}[H]
	\begin{tabular*}{\linewidth}{c @{\extracolsep{\fill}} cccc}
		\hline
		Data   \hspace{0.15cm} & $H_0$   & $\Omega_{m_0}$   & $\nu$ \\[2pt]
		\hline
		Data1  \hspace{0.15cm} & $70.5253\pm0.1345$  & $0.2628\pm0.0049$  & $0.0054\pm0.0011$ \\[2pt]
		\hline
		Data2  \hspace{0.15cm} & $68.8835\pm0.1290$  & $0.2679\pm0.0044$  & $0.0046\pm0.0011$ \\[2pt] 
		\hline
	\end{tabular*}
\caption{\label{tab:table2}The most probable values of the model parameters ($H_0$, $\Omega_{m_0}$ and $\nu$) for Data1 and Data2 are given in the table.}
\end{table}
\begin{table}[H]
	\begin{tabular*}{\linewidth}{c @{\extracolsep{\fill}} ccccc}
		\hline
		\vspace{0.1cm}
		Data  \hspace{0.15cm} & $\Zstroke(M_{RVM})$   & $\Zstroke(M_{\Lambda CDM})$   & $B_{ij}=\frac{\Zstroke(M_{RVM})}{\Zstroke(M_{\Lambda CDM})}$\vspace{0.1cm} \\
		\hline
		\vspace{0.1cm}
		Data1  \hspace{0.15cm} & $8.7657\times10^{-232}$  & $5.4089\times10^{-230}$  & $0.0162$ \vspace{0.1cm}\\ 
		\hline
		Data2  \hspace{0.15cm} & $1.1432\times10^{-236}$  & $1.2412\times10^{-235}$  & $0.0921$ \vspace{0.1cm}\\ 
		\hline
	\end{tabular*}
\caption{\label{tab:table3}The Bayesian evidences of the running vacuum model, $\Lambda$CDM model and the bayes factor for Data1 and Data2 are given in the table.}
\end{table}
The same procedure was repeated for $\Omega_{m_0}$ and $H_0$. The marginal PDF obtained for the combined data sets Pantheon+CMB+BAO (data1) and Pantheon+CMB+BAO+Hubble data (data2) are shown in above figure \ref{fig:nu1}--\ref{fig:om2}.

The Bayes factor of the running vacuum model compared to the standard $\Lambda$CDM model ($M_{\Lambda CDM}$) is obtained using the equation (\ref{eqn:p02}), which are summarized in Table \ref{tab:table3}.
The Bayes factor of the running vacuum model concerning $\Lambda$CDM model for the data combination Pantheon+CMB+BAO and Pantheon+CMB+BAO are obtained as 0.01620 and 0.09209, respectively. The Bayes factors, in both cases, are in the range, $B_{ij}<1$, according to Jeffreys scale, the evidence of running vacuum model against the $\Lambda$CDM model is weak for both data combinations.  

\section{Conclusions}
The $\Lambda$CDM model is regarded as the concordance model of cosmology due to its potential ability to explain the current observational data. Nevertheless, the model has some downsides, for instance, the vast discrepancy between the observed value of the cosmological constant and its forecasted value from quantum field theory in curved space-time. The modified gravity theories are considered as an alternative approach to tackle this problem and also explain the evolutionary history of the Universe; the running vacuum model is one of them which considers dynamical vacuum energy density as a measure of the cosmological constant. In this model, the vacuum energy density has gained its running status through the Hubble parameter. More precisely, the vacuum energy density is proportional to $H^2$.

In this work, we have adopted Bayesian inference to extract the model parameters testest the relative significance of running vacuum model against the $\Lambda$CDM model. The marginal PDF of the model parameters are well fitted with the Gaussian function. Its mean is the most probable value, and the standard deviation turns out to be its error.   We have obtained the likelihood of our model and $\Lambda$CDM model using different data combinations. The Bayes factor is the ratio of the likelihood of the running vacuum model and $\Lambda$CDM model. The Bayes factor obtained is much less than one for both data combinations. According to Jeffrey's scale of Bayesian inference, the evidence of the $\Lambda$CDM model is strong against the running vacuum model.
\section*{Acknowledgements}

One of the authors Sarath N is thankful to UGC, Govt. of India, for providing financial support through Junior Research Fellowship, Mr. Manosh T.M for suggestions on the manuscript and Subin P Surendran for suggestions on programming.
\section*{Data Availability}
The pantheon data underlying this article are available at \url{https://github.com/dscolnic/Pantheon.git}.
\bibliography{references}
\bibliographystyle{ieeetr}

\end{document}